\documentclass[a4paper,twoside]{article}
\newcommand{\affil}[1]{$^{\rm #1}$}
\usepackage[authoryear]{natbib}
\bibpunct{(}{)}{;}{a}{}{,}
\usepackage{graphicx}
\date{} 
\title{\large\bf\flushleft The Correlated Multi-color Optical Variations of BL Lac Object S5 0716+714}
\author{\parbox{\textwidth}{\flushleft
\vspace{-0.5cm}
{\it Bingkai Zhang\affil{A*}, Benzhong Dai\affil{B}, Li
Zhang\affil{B,C}, Jiali
Liu\affil{D}, and Zhen Cao\affil{D}}\\
\vspace{0.4cm}
{\small \affil{A}\,Department of Physics, Fuyang Normal University, Fuyang, P. R. China, 236041}\\
{\small \affil{B}\,Department of Physics, Yunnan University, Kunming, P. R. China, 650091}\\
{\small \affil{C}\,National Astronomical Observatories/Yunnan
Observatory, Chinese Academy of Sciences, P.O. Box 110, Kunming, P.
R. China, 650011}\\
{\small \affil{D}\,The Key Laboratory of Particle and Astrophysics,
  Institute of High Energy Physics, Chinese Academy of  Sciences, Beijing, P. R. China,
  100049}\\
{\small \affil{*}\,Email: zhangbk@mail.ihep.ac.cn}}}
\begin{document}
\twocolumn[
\begin{changemargin}{.8cm}{.5cm}
\begin{minipage}{.9\textwidth}
\vspace{-1cm}
\maketitle
%
%
\small{\bf Abstract:} S5 0716+714 is a well-studied BL Lac object in
the sky. Verifying the existence of correlations among the flux
variations in different bands serves as an important tool to
investigate the emission processes. To examine the possible
existence of a lag between variations in different optical bands on
this source, we employ a discrete correlation function (DCF)
analysis on the light curves. In order to obtain statistically
meaningful values for the cross-correlation time lags and their
related uncertainties, we perform Monte Carlo simulations called
``flux redistribution/random subset selection'' (FR/RSS). Our
analysis confirms that the variations in different optical light
curves are strongly correlated. The time lags show a hint of the
variations in high frequency band leading those in low frequency
band of the order of a few minutes.


\medskip
\medskip
\end{minipage}
\end{changemargin}
]
\small

\section{Introduction}
Blazars include BL Lacertae (BL Lac) objects and flat spectrum radio
quasars (FSRQs). They are the most extreme class of active galactic
nuclei (AGNs) and exhibit strong variability at all wavelengths of
the whole electron-magnetic (EM) spectrum, strong polarization from
radio to optical wavelengths, and are usually core dominated radio
structures. These extreme properties are generally interpreted by
many authors as a consequence of non-thermal emission from a
relativistic jet oriented close to the line of sight
\citep{bland79,urry95}. Blazars vary on the diverse time scales
(Gupta et al., 2008 and references therein). Variability has been
one of the most powerful tools in revealing the nature of blazars.
Understanding variation behaviour is one of the major issues of AGNs
studies. The variability lags between different energy bands provide
very important constrains for the interpretation of the emission
components. In recent years, correlations of the variability in
different energy regions have been widely studied
\citep{raite01,raite08,dai06,marsh08,areva08,chatt08,villa09,bonni09}.

The radio source S5 0716+714 is one of the brightest and
best-studied BL Lacertae objects. It was discovered in 1979
\citep{kuhr81}, and was classified as a BL Lac object because of its
featureless spectrum and its strong optical polarization
\citep{bierm81}. By optical imaging of the underlying galaxy, its
redshift of z = 0.31$\pm$0.08 was derived recently \citep{nilss08}.
It is a highly variable BL Lac object in the whole EM spectrum on
diverse time scales
\citep{heidt96,ghise97,villa00,raite03,pian05,bach05,bach06,
nesci05,ostor06,monta06,fosch06,wu07,wagne96,stali06}. The
correlations and time lags between different energy bands in this
source have been studied by some authors. A correlation was claimed
for a selected range of data at 6 and 3 cm wavelengths, and optical
wavelengths by \cite{quirr91}. But 2 cm data from the same epoch
published later by \cite{quirr00} does not show evidence for
correlated variability, casting doubt on the reliability of the
claimed correlation between radio and optical bands \citep{bigna03}.
A good correlation was noticed between the light curves in the
different passbands \citep{sagar99}. An upper limit to the possible
delay between $B$ and $I$ variations (10 minutes) was determined by
\cite{villa00}. With the data taken on January 8, 1995,
\cite{qian00} found an upper limit of the time lag of 0.0041 days
between variations in the $V$ and $I$ bands. \cite{raite03} found
that the variations between different radio bands are very
correlated, and the flux variations at the lower radio frequencies
are delayed with respect to those at the higher frequencies. They
also found weak correlations
 between the optical and radio emissions. \cite{stali06} analyzed two night
 data, found that $V$ and $R$ are correlated with a small time lag (about 6 and 13
minutes, respectively, for April 7 and 14, 1996) and the variation
at $V$ leading that at $R$ on both nights. \cite{chen08} studied the
gamma-optical correlation. Their result suggested a possible delay
in the gamma-ray flux variations with respect to optical ($R$ band)
variations of the order of 1 day. \cite{fuhrm08} confirmed the
existence of a significant correlation across all their observed
radio-bands. The time delay between the two most separated bands
($\lambda _{3mm}$ and $\lambda _{60mm}$) is about 2.5 days. It must
be noted that the time delay of 2.5 days is determined from an
almost monotonic increase in flux density observed over a time range
of 11 days.

The determination of time lag can be used to study the geometry,
kinematics and physical conditions in the inner regions of AGNs.
This is attributed to light travel-time effect \citep{villa00}. It
is very important to search for detectable delays between optical
bands themselves, even if these can be expected to be very small, if
any. Our goals in this paper are to investigate the correlations
between different optical light curves. The paper is arranged as
follows: in Sect. 2, the light curves are presented; then in Sect.
3, the correlations and the time lags between different optical
passbands are presented; after this, the discussion and conclusions
are given in Sect. 4.

\section{Light Curves}
S5 0716+714 is a well-monitored object. It has been observed in
various multi-frequency campaigns
\citep{wagne96,sagar99,villa00,villa08,stali06}. In addition,
\cite{ghise97} presented the results of their optical observations
between Nov. 15, 1994 and Apr. 30, 1995. \cite{qian02} published
$\emph{BVRI}$ band light curves from 1994 to 2000. \cite{raite03}
presented the largest optical database from 1994 to 2001.
\cite{xie04} also reported their monitoring result on this source.
\cite{monta06} introduced a large set of observational data
containing 10,675 photometric points. \cite{gu06} extensively
monitored this source in 2003 and 2004. \cite{zhang08} presented the
optical photometries ($\emph{BVRI}$) of the source from 2001
February to 2006 April.

 To search for the lags between different optical bands, one needs
well-sampled and high-quality datasets. Examining the datasets
mentioned above, we find the optical photometry presented by
\cite{gu06} provided a possibility to determine the time lag. The
source was observed quasi-simultaneously in four passbands and the
overall observations span about 140 days. Each optical light curve
in the $\emph{BVRI}$ bands is well-sampled. Furthermore, the source
exhibited strong variability during the observations.

\begin{figure}[ht]
\begin{center}
\includegraphics{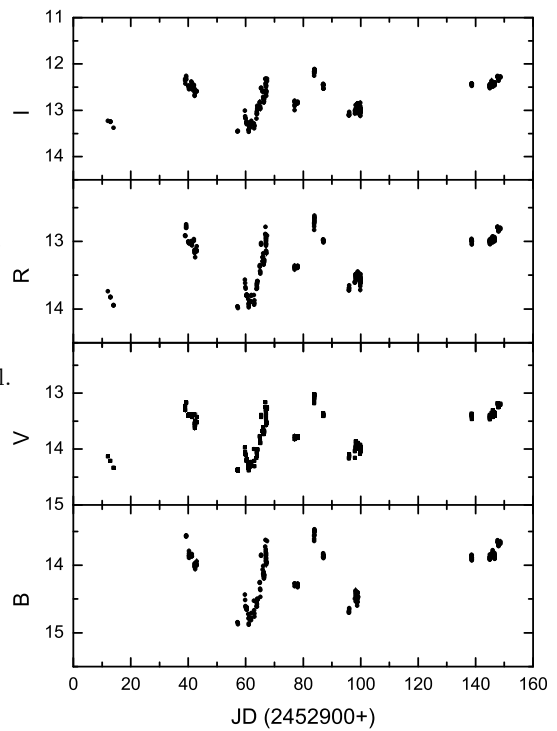}
\end{center}
\caption{Light curves of S5 0716+714 in the $BVRI$ bands.}
\label{fig:light-curves}
\end{figure}

\begin{figure}[ht]
\begin{center}
\includegraphics{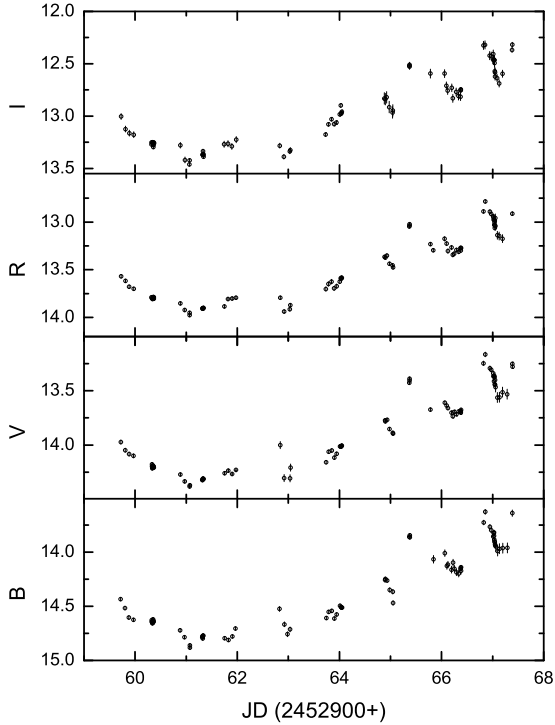}
\end{center}
\caption{Light curves of S5 0716+714 in the $BVRI$ bands during JD
2,452,958 $-$ 2,452,970.} \label{fig:sub-light-curves}
\end{figure}

The detailed statistics of these data are listed in
Table~\ref{tab:BVRI}. The first column represents the bands, the
second represents the numbers, the third represents the mean values,
the fourth represents the standard deviations, the fifth represents
the largest variations and the last represents the median interval
of the data points. The light curves of different passbands are
displayed in Fig. $\ref{fig:light-curves}$. They are very similar
but different amplitudes. In the light curves, there are some
significant substructures. Because of the large time scale, most
independent variations cannot be seen. To illustrate the fine
structures of variations, Fig. $\ref{fig:sub-light-curves}$ shows
the light curves during JD 2,452,958-2,452,970. It is clear that the
source does not undergo a monotonic increase between JD 2,452,958
and 2,452,970. The data contains many more variations than those
indicated by Fig. \ref{fig:light-curves}. The source exhibits
independent variation almost every night.

\begin{table}[ht]
\centering \caption[]{ Statistics of $BVRI$-band Data of BL Lac
Object S5 0716+714}\label{tab:BVRI}
 \begin{tabular}{@{}cccccc@{}}
  \hline
$Band$   & $N$ &  $Mean$ & $\sigma$ & $\Delta $ & Interval\\
         &     &   (mag.)&  (mag.)  &  (mag.) &  (min.)\\
  \hline
    $B$       & 239  & 14.13 & 0.38 & 1.42 & 11.2    \\ 
    $V$       & 287  & 13.69 & 0.37 & 1.37 & 11.1    \\ 
    $R$       & 301  & 13.28 & 0.36 & 1.37 & 11.1   \\ 
    $I$       & 292  & 12.75 & 0.35 & 1.36 & 11.1    \\
  \hline
\end{tabular}

\end{table}

\begin{figure}[ht]
\begin{center}
\includegraphics[width=8cm]{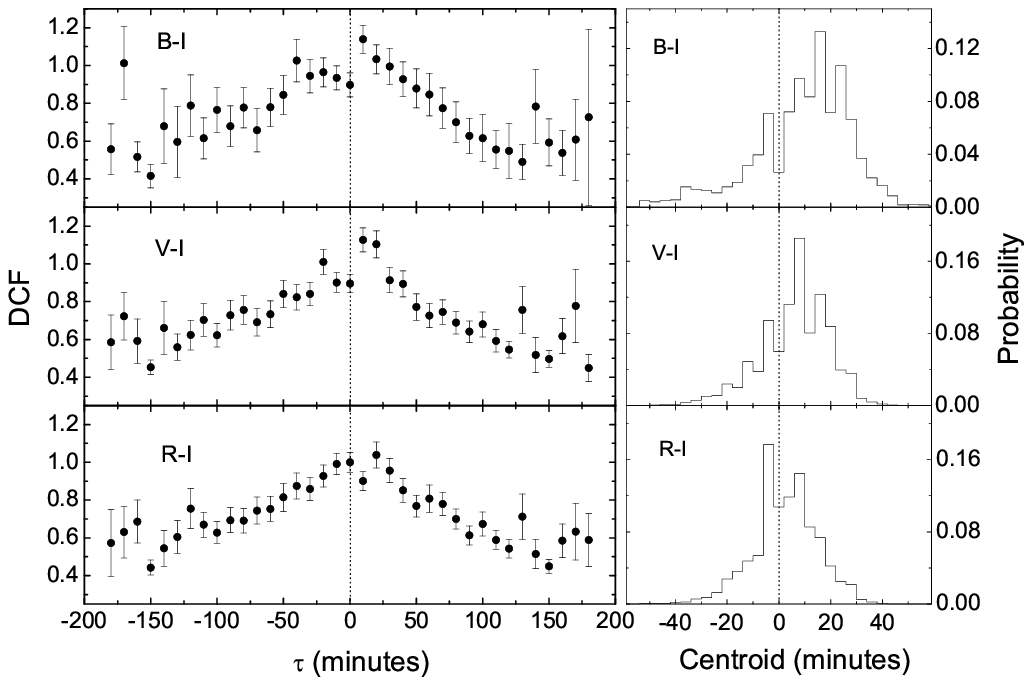}
\end{center}
\caption{Left: DCFs between $B$ and $I$, $V$ and $I$, $R$ and $I$,
respectively. Right: Normalized CCPDs relative to the central peak
obtained by running 8000 FR/RSS Monte Carlo simulations. Dashed
vertical lines are drawn to guide the eye.} \label{fig:dcf1}
\end{figure}

\begin{figure}[ht]
\begin{center}
\includegraphics[width=8cm]{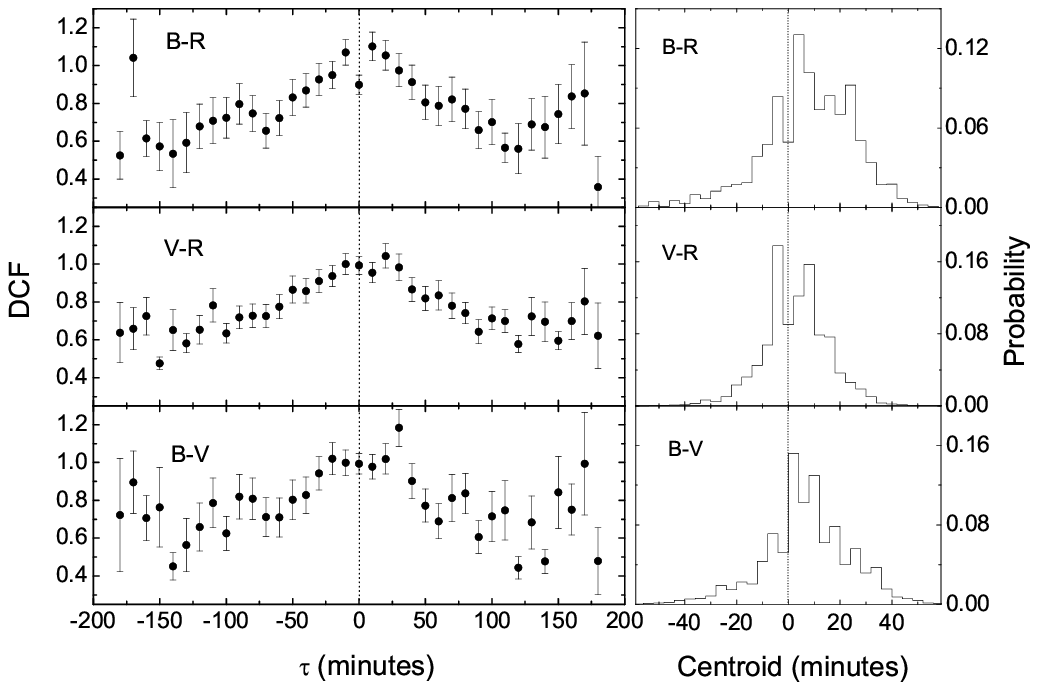}
\end{center}
\caption{Left: DCFs between $B$ and $R$, $V$ and $R$, $B$ and $V$,
respectively. Right: Normalized CCPDs relative to the central peak
obtained by running 8000 FR/RSS Monte Carlo simulations. Dashed
vertical lines are drawn to guide the eye.} \label{fig:dcf2}
\end{figure}

\begin{figure}[ht]
\begin{center}
\includegraphics[width=8cm]{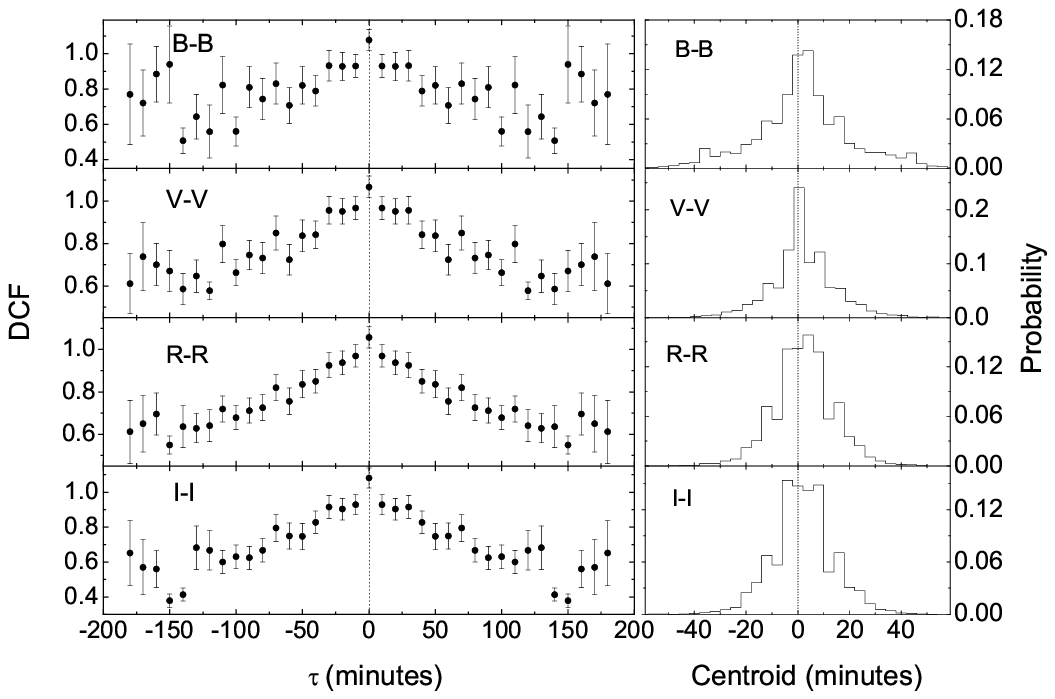}
\end{center}
\caption{Left: Auto-DCFs of $B-B$, $V-V$, $R-R$ and $I-I$,
respectively. Right: Normalized CCPDs relative to the central peak
obtained by running 8000 FR/RSS Monte Carlo simulations. Dashed
vertical lines are drawn to guide the eye.} \label{fig:dcf3}
\end{figure}

\section{Correlation Analysis}
\subsection{Discrete Correlation Function \\Method}
The discrete correlation function (DCF) was introduced by
\cite{edels88}. It is a useful method of measuring correlation, and
it does not require interpolating in the temporal domain. This
method can not only provide the correlation of two series of
unevenly sampled variability data with the time lag, but also give
the evidence of periodicity that lies in a single temporal data set.
Its other advantages are that it uses all the data points available
and calculates a meaningful error estimate. Our first step is to
calculate the set of un-binned discrete correlations (UDCF) between
each data point in the two data streams. It is defined as follows:
\begin{equation}
UDCF_{ij}=\frac{(a_i-\overline{a})\times(b_j-\overline{b})}{\sqrt{\sigma_a^2\times\sigma_b^2}},
\end{equation}

where $a_i$  and $b_j$  are points of the data sets $a$ and $b$,
$\overline{a}$ and $\overline{b}$ are the means of the data sets $a$
and $b$, $\sigma_a$ and $\sigma_b$ are the standard deviations of
each data set. Each of UDCF is associated with the pair-wise lag
$\Delta t_{ij}=t_j-t_i$. Then we average over the $M$ pairs for
which $\tau-\Delta \tau/2\leq\Delta t_{ij}<\tau+\Delta \tau/2$, and
obtain DCF:

\begin{equation}
DCF(\tau)=\frac{1}{M}\Sigma UDCF_{ij}(\tau),
\end{equation}

where $M$ is the number of pairs in the bin. When $a = b$, the
autocorrelation DCF is produced, and when $a \neq b$, the
cross-correlation DCF is measured. In most case, the evident peak in
the cross-correlation function means a strong correlation between
two data series, and the peak in the autocorrelation DCF implies a
strong period in the data set. The standard error for each bin is
defined as

\begin{equation}
\sigma(\tau)=\frac{1}{M-1}\{{\Sigma{[UDCF_{ij}-DCF(\tau)]}^2}\}^{1/2}.
\end{equation}

In order to obtain statistically meaningful values for the
cross-correlation time lags and their related uncertainties, it is
common usage to calculate the centroid $\tau_{c}$ of the DCF, given
by

\begin{equation}
\tau_{c}=\frac{\sum_{i}\tau_{i}DCF_{i}}{\sum_{i}DCF_{i}},
\end{equation}

where sums run over the points which have a DCF value close to the
peak one ($DCF_{i}>0.8DCF_{peak}$), then perform Monte Carlo
simulations known as ``flux redistribution/random subset selection''
(FR/RSS) described in detail by \cite{peter98} and \cite{raite03}.
Random subsets of the two datasets to be correlated are selected,
redundant points are discarded, and random gaussian deviates
constrained by the flux errors are added to the fluxes;
 Thus,
the influence of both uneven sampling and flux density errors is
taken into account. In each simulation, the two subsets are then
cross-correlated and the centroid $\tau_{c}$ of the DCF peak is
determined. After a large number of simulations (generally
500$\sim$2000), the cross-correlation peak (actually, the centroid)
distribution (CCPD) is obtained. As the measures of the time lag and
its uncertainties, $\tau_{median}$ and $\pm\Delta\tau_{68}$ can be
computed directly from the CCPD, where $\pm\Delta\tau_{68}$
corresponds to $1\sigma$ errors for a normal distribution
\citep{peter98}.

\subsection{Result}
According to the above method, we calculate the discrete correlation
function (DCF) between different optical bands to search for
correlations and possible time lags. Between the $\emph{B}$ and $I$
band light curves, the median interval time is 6.3 minutes. The DCF
result computed with a bin size of 10 minutes is shown in the top
left panel of Fig \ref{fig:dcf1}. The curve of DCF has an obvious
maximum at the position of 10 minutes. The maximum value of DCF is
1.1 $\pm$ 0.1. The centroid $\tau_{c}$ corresponding this peak is
0.1 minutes. To give a statistic reliable result, we perform Monte
Carlo simulations. Because the possible delay is of the same
magnitude as the DCF bin size and of the light curve time
resolution, the effect of bin size cannot be ignored. To eliminate
the effect caused by the choice of the bin size, we calculate the
DCF with different bin sizes from 8 to 15 minutes. The lack of a
better sampling does not allow us to improve the bin size resolution
without increasing spurious effects too much. We perform 1000 Monte
Carlo simulations for each bin size, thus 8000 Monte Carlo
simulations are performed. The CCPD is obtained and plotted in the
top right panel of Fig \ref{fig:dcf1}. From this, the time lag and
the uncertainties of $10.2^{-20.7}_{+13.8}$ minutes are derived. In
addition, the mean $DCF_{peak}$ of 1.2 $\pm$ 0.1 is achieved by 8000
Monte simulations. This means that the variation of $B$ band is
strongly correlated with that of $I$ band with leading about
$10.2^{-20.7}_{+13.8}$ minutes.

This procedure is applied to each two bands of $BVRI$.  The DCFs and
CCPDs are plotted in Figs. \ref{fig:dcf1} and \ref{fig:dcf2}. The
results are shown in Table \ref{tab:correlation}. The first column
gives the band, the second one gives the median time interval of two
bands, the third one gives the possible delay and the last one gives
the mean value of $DCF_{peak}$s. The results suggest the existence
of significant correlation among $BVRI$ band light curves and the
higher frequencies varying earlier.

\begin{table}[h]
\newcommand\T{\rule{-5pt}{2.6ex}}
\newcommand\B{\rule{0ex}{-5pt}}
\begin{center}
\caption{The Possible Time Lags Determined by
DCF}\label{tab:correlation}
\begin{tabular}{@{}ccccc@{}}
\hline
$Band$  \T & $Interval(min.)$ & $Lag(min.)$ & $DCF_{peak}$   \\[3pt]
  \hline
    $B-I$ \T    & 6.3 & $10.2^{-20.7}_{+13.8}$  & 1.2 $\pm$ 0.1      \\[3pt] 
    $V-I$       & 4.9 & $5.8^{-14.9}_{+11.4}$   & 1.1 $\pm$ 0.1      \\[3pt] 
    $R-I$       & 3.7 & $0.2^{-11.3}_{+12.4}$   & 1.0 $\pm$ 0.1     \\[3pt] 
    $B-R$       & 5.3 & $6.0^{-15.9}_{+16.9}$   & 1.1 $\pm$ 0.1     \\[3pt]
    $V-R$       & 4.3 & $0.3^{-10.6}_{+12.3}$   & 1.1 $\pm$ 0.1     \\[3pt]
    $B-V$      & 5.2 & $5.6^{-14.6}_{+16.8}$   & 1.1 $\pm$ 0.1    \\[3pt]
  \hline
\end{tabular}
\end{center}
\end{table}

\begin{table}[h]
\newcommand\T{\rule{-5pt}{2.6ex}}
\newcommand\B{\rule{0ex}{-5pt}}
\begin{center}
\caption{Test Results between Light Curves and Their Corresponding
Shifted Light Curves$^{*}$}\label{tab:test}
\begin{tabular}{@{}ccccc@{}}
\hline
$t$ \T  & $lag_{B}$         & $lag_{V}$         & $lag_{R}$         & $lag_{I}$  \\[3pt]
  \hline
    $1$ \T & $0.4^{-13.9}_{+14.4}$ & $0.7^{-11.5}_{+12.0}$ &  $0.6^{-11.7}_{+12.3}$ & $0.5^{-11.4}_{+12.9}$  \\[3pt] 
    $2$ & $3.0^{-13.3}_{+12.0}$ & $1.0^{-9.9}_{+11.4}$ &  $0.9^{-11.2}_{+12.5}$ & $0.6^{-11.4}_{+13.2}$   \\[3pt] 
    $3$ & $1.3^{-12.0}_{+14.9}$ & $3.9^{-11.1}_{+9.9}$ &  $4.0^{-12.3}_{+10.1}$ & $3.4^{-11.7}_{+11.2}$   \\[3pt] 
    $4$ & $4.5^{-15.2}_{+15.0}$ & $4.7^{-11.1}_{+11.4}$ &  $4.9^{-11.1}_{+11.6}$ & $4.6^{-10.9}_{+12.0}$  \\[3pt] 
    $5$ & $4.6^{-15.3}_{+14.3}$ & $4.6^{-10.9}_{+10.8}$ &  $4.9^{-11.3}_{+11.6}$ & $4.9^{-11.6}_{+10.6}$  \\[3pt] 
    $6$ & $5.7^{-12.8}_{+12.4}$ & $5.7^{-10.8}_{+10.0}$ &  $5.8^{-11.8}_{+11.8}$ & $5.8^{-11.9}_{+11.2}$  \\[3pt] 
    $7$ & $7.1^{-14.8}_{+14.4}$ & $7.0^{-11.7}_{+12.6}$ &  $7.3^{-12.7}_{+12.9}$ & $6.9^{-12.2}_{+12.9}$  \\[3pt] 
    $8$ & $8.0^{-15.5}_{+16.6}$ & $8.4^{-12.2}_{+12.4}$ &  $8.4^{-12.7}_{+12.4}$ & $7.8^{-12.3}_{+13.3}$  \\[3pt] 
    $9$ & $9.3^{-12.3}_{+14.6}$ & $9.5^{-10.0}_{+11.1}$ &  $9.2^{-9.7}_{+12.4}$ & $8.8^{-11.7}_{+12.5}$   \\[3pt] 
   $10$ \B & $10.6^{-15.7}_{+12.6}$ & $10.2^{-10.1}_{+10.4}$ &  $10.3^{-10.2}_{+11.5}$ & $10.0^{-9.9}_{+12.0}$\\[3pt] 
  \hline
\end{tabular}
\medskip\\
$^*$The mean of each column can be seen in the text, and all the above are in units of minutes.\\
\end{center}
\end{table}

\section{Discussion and Conclusions}
It is well known that the light curves between variant wave bands
may have a short time lag. According to the inhomogeneous jet
models, time delays are expected between the emission in different
energy bands, as plasma disturbances propagate downstream
\citep{georg98}. Multi-wavelength monitoring of blazars shows that
flares usually begin at high frequencies and then propagate to lower
frequencies, implying that high-frequency synchrotron emission
arises closer to the core than low-frequency synchrotron emission
does \citep{ulric97,marsc01}. High energy electrons emit synchrotron
radiation at high frequencies and then cool, emitting at
progressively lower frequencies and resulting in time lags between
high and low frequencies \citep{bai05}. The small time lags in
optical regimes may be result of very small frequency intervals, and
may indicate that the photons in these wavelengths should be
produced by the same physical process.

The above results have been obtained with a light-curve time
resolution of the same magnitude as the possible lag. To check the
method and the result mentioned above, we calculate the $B$-$B$ band
autocorrelation DCF (Fig. \ref{fig:dcf3}). The result of Monte Carlo
simulations suggests that the time delay is $-0.0^{-16.3}_{+15.5}$
minutes. For the $V$-$V$, $R$-$R$ and $I$-$I$ autocorrelations,
$\tau$s are $0.0^{-11.4}_{+11.6}$, $0.1^{-11.5}_{+12.0}$ and
$-0.0^{-11.5}_{+11.8}$ minutes(see Fig. \ref{fig:dcf3}). They are
all consistent with the expectation of zero lags. Furthermore, we
shift $BVRI$ band light curves by a time of $t_{shift}$ and generate
the corresponding artificial light curves of $B_{shift}$,
$V_{shift}$, $R_{shift}$ and $I_{shift}$. Using the same process, we
search for the time lag between the light curve and the shifted
light curve. The results are listed in Table \ref{tab:test}. The
first column is for the shifted time , the second to the last are
for the lags of $B$-$B_{shift}$, $V$-$V_{shift}$, $R$-$R_{shift}$
and $I$-$I_{shift}$ determined by the method introduced above. From
the results listed in Table \ref{tab:test}, one can see that the
lags between two correlated light curves can be well determined. In
addition to shifting the light curves to calculate the displaced
Auto-DCF peaks, we shift the light curves with respect to those at
each other wavelength by the time lags determined. The resulting
DCFs are more consistent with zero lag. So, the time lags between
different optical $BVRI$ band light curves are more reliable.

The time lag of 10.2 minutes between variations of $B$ and $I$ band
 is consistent with the result found by \cite{villa00}. They derived
an upper limit of 10 minutes between $B$ and $I$ variations. The
5.8-minute lag between the band $V$ and $I$ is in good agreement
with the 0.0041-day lag determined by \cite{qian00}. For the $V$ and
$I$ band variations, \cite{stali06} reported time lags of about 6
and 13 minutes on two individual nights. Between the $V^{'}$ and
$R^{'}$ band variations, \cite{wu07} obtained the lags of 2.34 $\pm$
5.25 minutes on JD 2,453,737 and -0.01 $\pm$ 1.88 minutes on JD
2,453,742. Because the time lags are very short, shorter than their
typical sampling interval, they concluded that they did not detect
an apparent time lag. In this analysis, it also must be pointed out
that the time lags between the different optical bands are of the
same orders of the light-curve time resolution, and the
uncertainties are larger than the lags. So, the lags should be
treated with caution, and much more significant results can be
achieved only with much denser monitoring data. Two ways were
suggested by \cite{villa00} to improve the time resolution of the
light curves, one way is to enlarge the telescope, another is to
synchronized use two or more telescopes.

In summary, the variations between different optical bands have been
analyzed and the time lags and their uncertainties are convincingly
determined. The results suggest that the variations are correlated
very strongly.  Considering the errors, every one of the values
agrees with zero-lag well. While looking at the whole set of lags,
it is possible to say that there is a hint of the variations of high
frequency bands leading those of low frequency bands in the order of
a few minutes.

\section*{Acknowledgments} 
 We express our thanks to the people helping with this work, and
acknowledge the valuable suggestions from the peer reviewers. This
work is supported by the Research Foundation of Education Department
of Anhui Province, China(KJ2010B159), National Natural Science
Foundation of China (10975145) and Fuyang Normal University, also
partly by Natural Science Foundation of Yunnan Province (2007A026M)
and the Innovation fund (U-526) of the Institute of High Energy
Physics.


\end{document}